\documentclass{aa}

\usepackage{newtxtext,newtxmath}

\usepackage[T1]{fontenc}
\usepackage{ae,aecompl}

\usepackage{natbib}
\bibpunct{(}{)}{;}{a}{}{,} 

\usepackage{graphicx}	
\usepackage{amsmath}	
\usepackage{amssymb}	
\usepackage{xcolor}
\usepackage[normalem]{ulem}
\usepackage{subcaption}

\providecommand{\gaia}{\textit{Gaia }}
\providecommand{\gaianospace}{\textit{Gaia}}
\providecommand{\kms}{km s$^{-1}$ }

\begin{document}

\title{Application of a Neural Network classifier for the generation of clean Small Magellanic Cloud stellar samples\thanks{The SMC / MW classification probability of each object will be made available in electronic form at the CDS via anonymous ftp to cdsarc.u-strasbg.fr (130.79.128.5) or via http://cdsweb.u-strasbg.fr/cgi-bin/qcat?J/A+A/}}

\titlerunning{Small Magellanic Cloud classifier for the generation of clean stellar samples}

\author{Ó. Jiménez-Arranz\inst{1,2,3}
   \and M. Romero-Gómez\inst{1,2,3}
   \and X. Luri\inst{1,2,3}
   \and E. Masana\inst{1,2,3}
}

\institute{{Departament de Física Quàntica i Astrofísica (FQA), Universitat de Barcelona (UB), C Martí i Franquès, 1, 08028 Barcelona, Spain}
\and
{Institut de Ciències del Cosmos (ICCUB), Universitat de Barcelona, Martí i Franquès 1, 08028 Barcelona, Spain}
\and
{Institut d’Estudis Espacials de Catalunya (IEEC), C Gran Capità, 2-4, 08034 Barcelona, Spain}
}

\date{Received <date> / Accepted <date>}

\abstract 
{Previous attempts to separate Small Magellanic Cloud (SMC) stars from the Milky Way (MW) foreground stars are based only on the proper motions of the stars.}
{In this paper we develop a statistical classification technique to effectively separate the SMC stars from the MW stars using a wider set of \gaia data. We aim to reduce the possible contamination from MW stars compared to previous strategies.} 
{The new strategy is based on neural network classifier, applied to the bulk of the \gaia DR3 data. We produce three samples of stars flagged as SMC members, with varying levels of completeness and purity, obtained by application of this classifier. Using different test samples we validate these classification results and we compare them with the results of the selection technique employed in the \gaia Collaboration papers, which was based solely on the proper motions.}
{The contamination of MW in each of the three SMC samples is estimated to be in the $10-40\%$; the ``best case'' in this range is obtained for bright stars ($G > 16$), which belong to the $V_{los}$ sub-samples, and the ``worst case'' for the full SMC sample determined by using very stringent criteria based on StarHorse distances. A further check based on the comparison with a nearby area with uniform sky density indicates that the global contamination in our samples is probably close to the low end of the range, around $10\%$.}
{We provide three selections of SMC star samples with different degrees of purity and completeness, for which we estimate a low contamination level and have successfully validated using SMC RR Lyrae, SMC Cepheids and SMC/MW StarHorse samples.}



\maketitle

\section{Introduction}

This paper is a follow-up of \cite{jimenez-arranz} (hereafter, J22). In that paper the authors analyzed the kinematics of the Large Magellanic Cloud (LMC) using the \gaia DR3 data; the analysis required a reliable separation of LMC and foreground (Milky Way) stars in the dataset; for this purpose a classification method based on a Neural Network was developed, tested and applied. The result was a series of datasets providing a reliable selection of LMC objects, published through the Centre de Donn\'ees de Strasbourg for public use.

In this paper we extend the application of this methodology to the Small Magellanic Cloud (SMC), in order to obtain similarly reliable datasets for the study of this object, and we also make them public for general use.

The paper is organized as follows. In Section~\ref{sec:data} we describe the \gaia base sample and the training sample. In Section~\ref{sec:classification} we explain how we train the classifier and apply it to the \gaia base sample. We also compare the different datasets obtained. Next, in Section~\ref{sec:NNvalidation}, we validate the data sets with external data, such as Cepheids \citep{ripepi17}, RRLyrae \citep{muraveva18} and StarHorse \citep{anders22}. Finally, we give our conclusions in Section~\ref{sec:conc}.

\section{Data selection}
\label{sec:data}

In this section we introduce the samples used in this paper. First, we characterise the \gaianospace~DR3 base sample \citep{gaiadr3} with stars selected around the SMC center. The contamination of foreground MW stars in this sample is non-negligible. One may think on distinguishing SMC and MW through their distances, however, due to the large uncertainties in the parallax-based distances at SMC \citep{edr3_astrometric} it is not possible and would only be effective when subtracting bright MW stars. Second, we characterise the \gaia training sample we use to train the machine learning classifier (Neural Network) to discriminate SMC stars from MW foreground stars. This training sample intends to mimic the full dataset available in the \gaia catalogue.

\subsection{\gaia base sample}
\label{sec:gaia_base_sample}

The \gaia base sample was obtained using a selection from the \texttt{gaia\_source} table in \gaia DR3 with a $10^{\circ}$ radius around the SMC centre defined as $(\alpha, \delta) = (12.80^{\circ}$, $-73.15^{\circ})$ \citep{cioni2000a} and a limiting $G$ magnitude of $20.5$. We only kept the stars with parallax and integrated photometry information, since they are used in the SMC/MW classification. This selection can be reproduced using the following ADQL query in the \gaia archive:

\begin{verbatim}
SELECT * FROM gaiadr3.gaia_source as g
WHERE 1=CONTAINS(POINT('ICRS',g.ra,g.dec),
CIRCLE('ICRS',12.80,-73.15,10))
AND g.parallax IS NOT NULL
AND g.phot_g_mean_mag IS NOT NULL
AND g.phot_bp_mean_mag IS NOT NULL
AND g.phot_rp_mean_mag IS NOT NULL
AND g.phot_g_mean_mag < 20.5
\end{verbatim}

The resulting base sample contains a total of 4~047~225 objects.

\subsection{\gaia training sample}

As in J22, we use GOG \citep{luri14} to produce a training data set of similar characteristics to the base sample. We select particles within $10^\circ$ around the SMC centre. We make it compatible with recent estimations of the mean distance and systemic motion obtained from EDR3 data: a distance of $62.8$ kpc \citep{cioni2000b} and a systemic motion of $\mu_{\alpha*} = 1.858$ $\mathrm{mas}\,\mathrm{yr}^{-1}$, $\mu_{\delta} = 0.385$  $\mathrm{mas}\,\mathrm{yr}^{-1}$ as inferred in the linear fit (Table 4) to the proper motions in \citet{luri20} (hereafter, MC21).

\begin{figure*} 
    \centering
    \includegraphics[width=1\textwidth]{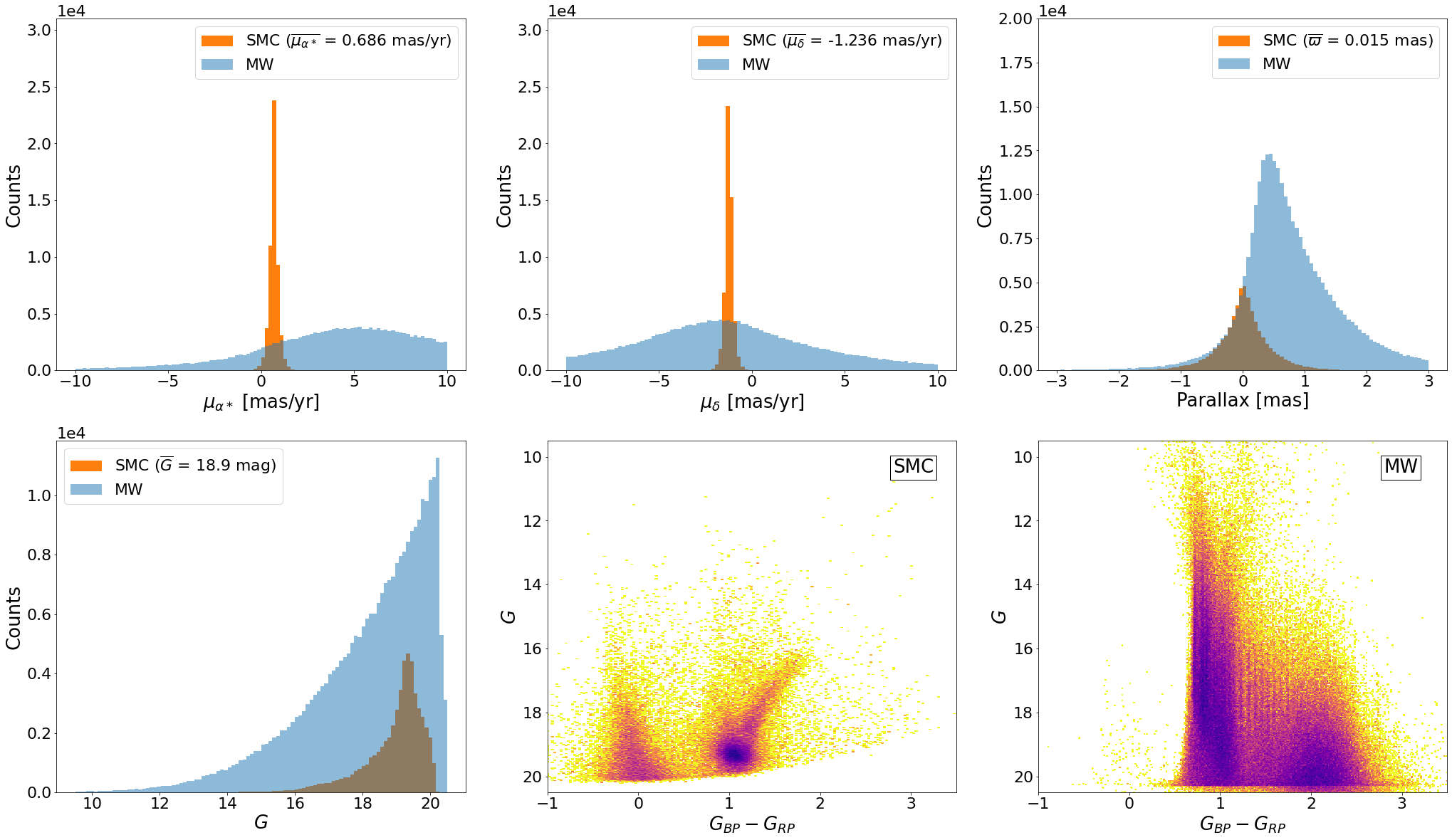}
    \caption{Characteristics of the GOG simulated samples, in orange and blue:\ SMC and the MW training samples, respectively. Top left and middle: Distribution of proper motions in right ascension and declination, respectively. Top right: Parallax distribution. Bottom left: Magnitude $G$ distribution of the simulated samples. Bottom middle and right: Colour-magnitude diagram of the SMC and MW, respectively. Colors represent relative stellar density, with darker colors meaning higher densities.}
    \label{fig:GOG}
\end{figure*}

The \gaia training sample is split into two labelled subsets one containing SMC and the other MW stars. The SMC simulation includes 54~109 sources, a smaller number of stars in comparison to what expected for the data. That is because the GOG simulator is based on a pre-defined catalogue of OGLE stars to provide real positions for the SMC stars \citep[see details in][]{luri14}. On the other hand, the MW simulation is based on a realistic galactic model which generates a number of stars that matches the observations. Similarly to the strategy used in J22, we compensate this unbalanced and unrealistic ratio between SMC and MW stars by retaining a random $20\%$ fraction of the MW simulation, obtaining 285~258 sources. In Figure~\ref{fig:GOG} both SMC and MW training subsets are characterised. 

\begin{figure*}
    \centering
    \includegraphics[width=1\textwidth]{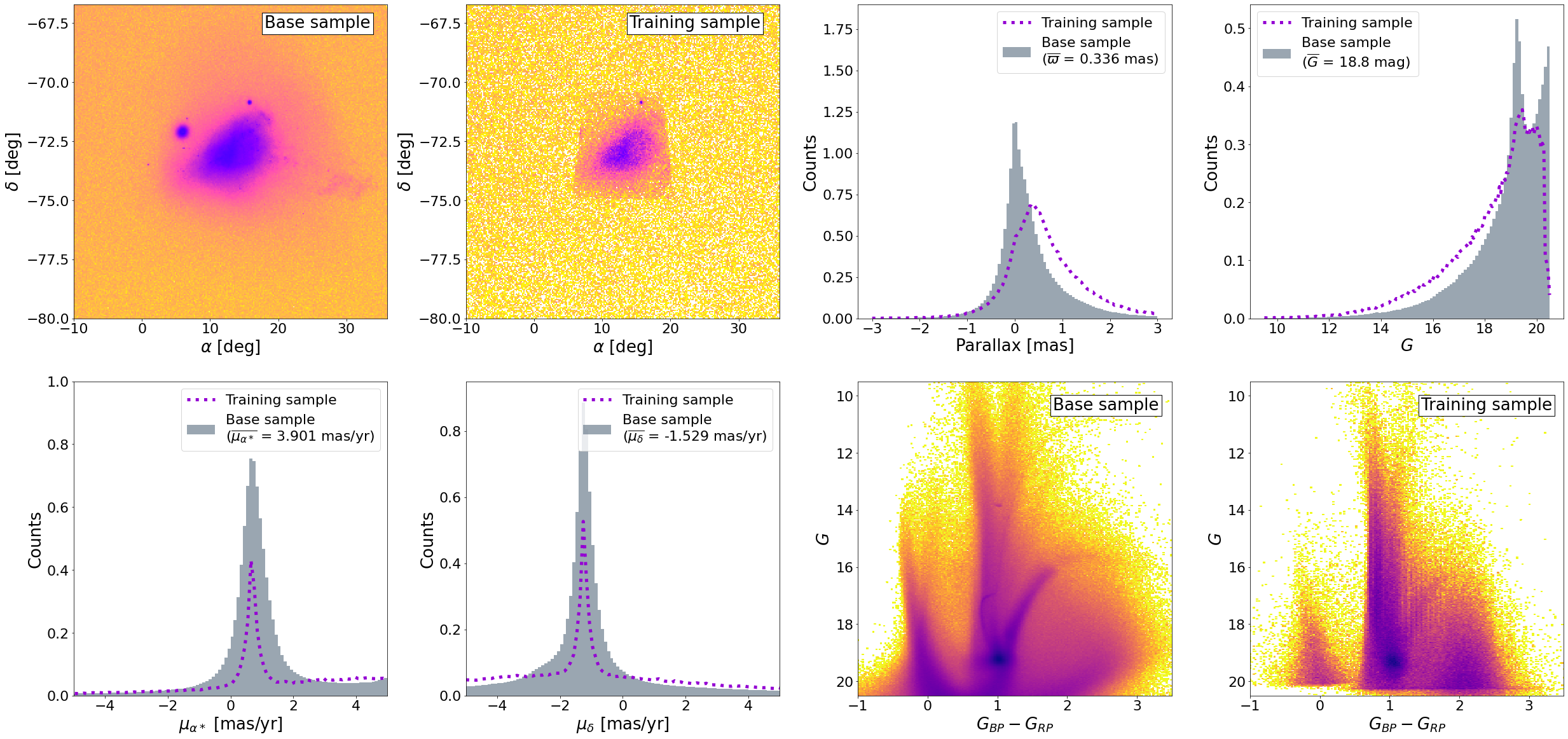}
    \caption{\gaia base and training samples comparison. Top from left to right: Density distribution in equatorial coordinates of the \gaia base and \gaia training samples in logarithmic scale, parallax, and G-magnitude distributions. Bottom from left to right: Proper motion distributions in right ascension and declination and colour-magnitude diagrams for the \gaia base and training samples. In the histograms, in gray we show the \gaia base sample, while in dotted purple we show the \gaia training sample. In the color-magnitude diagrams, colors represent relative stellar density with darker colors meaning higher densities.}
    \label{fig:SMCbasesample}
\end{figure*}

Our training sample is the result of combining these two simulations, which we contrast with the \gaia base sample in Figure ~\ref{fig:SMCbasesample}. These plots demonstrate that the \gaia training sample roughly matches the major characteristics of the \gaia base sample, but also highlights some of its limitations. For example, the colour-magnitude diagram (CMD) for the SMC simulation is not fully representative at the faintest magnitudes, with a lack of stars and an artificial cut line, and the distribution of the SMC stars in the sky forms a kind of square due to its origin based on an extraction from the OGLE catalogue. We will test their effectiveness using a number of validation samples to ensure that they are appropriate. 

\subsection{Proper motions-based classification}
\label{sec:PM}

To establish a baseline comparison with previous methods, we use the same selection based on the proper motions as in MC21. In short, the MW foreground contamination is minimized by computing the median proper motions of the SMC from a sample constrained to its very centre plus a cut in magnitude and parallax. We keep only stars whose proper motions obey the constraint of $\chi^2 < 9.21$, that is, an estimated 99\% confidence region (see details in Section~2.2 of MC21). The resulting sample (hereafter, PM selection) contains 1~720~856 objects\footnote{Note that the difference in the number of sources with the ones in MC21 comes from the different cut in radius, now being of $10^{\circ}$ instead of $11^{\circ}$.}.

\section{SMC/MW classification}
\label{sec:classification}

In this section we define an improved, more efficient and adjustable selection strategy to distinguish the SMC stars from the Milky Way foreground. Then, based on this classifier, we select three samples of candidate SMC stars with different degrees of completeness and purity.

\subsection{Training the classifier}

The sklearn Python package \citep{sklearn} was used to create a classifier. Using the \gaia data, this module includes a number of classifiers that can be used to differentiate the MW foreground objects from the SMC objects in our base sample using the training sample mentioned in the preceding section. We use position ($\alpha$, $\delta$), parallax and its uncertainty ($\varpi$, $\sigma_{\varpi}$), along with the proper motions and their uncertainties ($\mu_{\alpha*}$, $\mu_\delta$, $\sigma_{\mu_{\alpha*}}$, $\sigma_{\mu_\delta}$), and \gaia photometry ($G$, $G_{BP}$, $G_{RP}$).

As in J22, we select as classifier the Neural Network (NN). The NN has 11 input neurons, corresponding to the 11 \gaia parameters listed above; three-hidden-layers with six, three and two nodes, respectively; and a single output which gives for each object the probability $P$ of being a SMC star (or, conversely, the probability of not being a MW star). The object is very likely to belong to the SMC (MW) if the $P$ value is close to 1 (0). The activation function that we employed was the Rectified Linear Unit (ReLU). With a constant learning rate, stochastic gradient descent is used in our model to optimize the log-loss function. The strength of the L2 regularization term is 1e-5.\footnote{The corresponding author can be contacted if readers are interested in using the Neural Network developed in the paper.}

To train the algorithm, we used $60\%$ of the training sample, and the remaining $40\%$ was used for testing purposes. By creating the Receiver Operating Characteristic (ROC) curve and computing the Area Under the Curve (AUC), we assessed the classifier performance. One of the most crucial evaluation criteria for determining the effectiveness of any classification model is the ROC curve. Using various probability thresholds, it summarizes the trade-off between the true positive rate and false positive rate. Another useful tool for classifier evaluation is the AUC of the ROC curve. The larger the AUC, the better the classifier works. An excellent model has an AUC that is close to 1, indicating that it has a high level of separability. Having an AUC equal to 0.5 indicates that the model is incapable of classifying the data.

\begin{figure}
    \centering
    \includegraphics[width=0.50\textwidth]{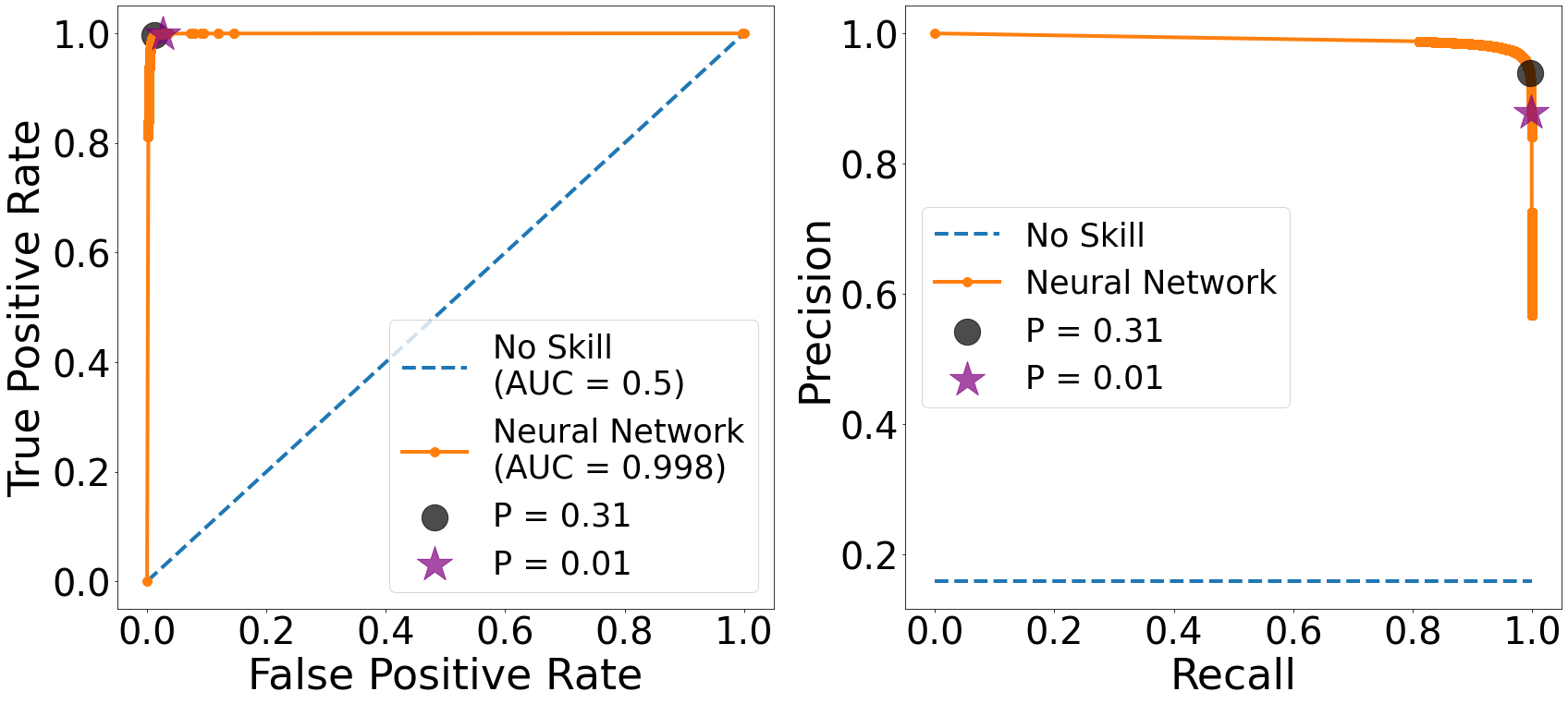}
    \caption{Evaluation metrics for the Neural Network classifier performance. Left: ROC curve.  Black dot is in the ``elbow'' of the ROC curve and it shows the best balance between completeness and purity. The purple star shows the completeness threshold. Right: Precision-Recall curve. In both cases, we compare our model (orange solid curve) with a classifier that has no class separation capacity (blue dashed curve).}
    \label{fig_roc_precission_recall}
\end{figure}

We provide the ROC curve of our NN classifier in the left panel of Figure~\ref{fig_roc_precission_recall}. We achieve an AUC of 0.998, indicating that our classifier accurately distinguishes between SMC and MW stars in the test sample. We show the Precision-Recall curve in the right panel of Figure~\ref{fig_roc_precission_recall}. When the classes are severely unbalanced, it is another helpful indicator to assess the output quality of the classifier. Both evaluation criteria display a nearly flawless classifier when applied to the training (simulated) data, however, same warnings regarding the classifier described in J22 apply here.

\subsection{Applying the classifier to the \gaia base data}\label{sec:NNapplication}

\begin{figure}
    \centering
    \includegraphics[width=0.50\textwidth]{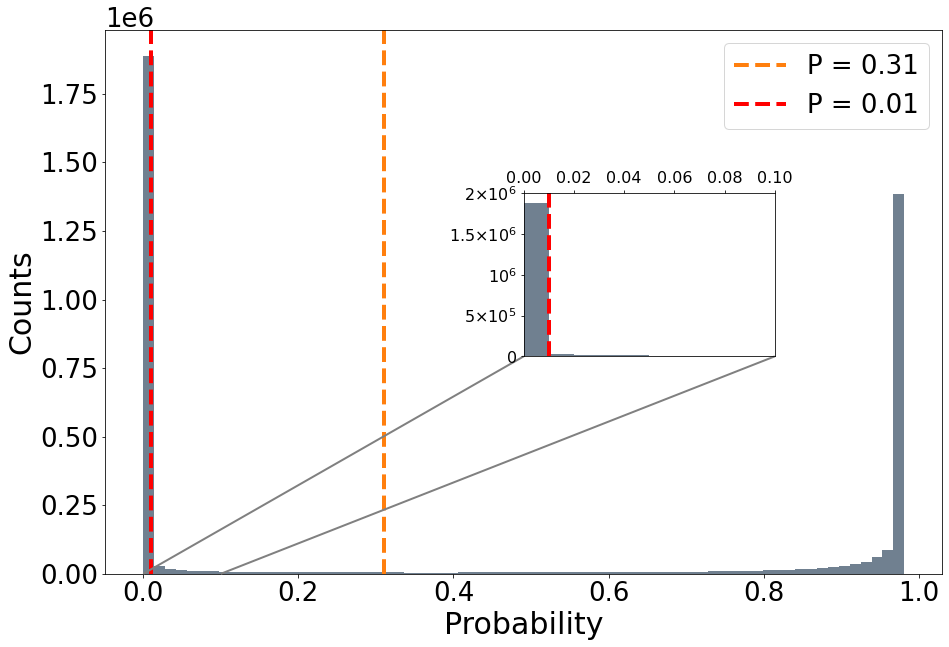}
    \caption{\gaia base sample's probability distribution for the NN classifier. A high likelihood of being an SMC (MW) star is indicated by a probability value close to 1 (0).}
    \label{fig_proba}
\end{figure}

After the NN has been trained, we use it to extract probabilities for each object in the \gaia base sample\footnote{By anonymous ftp to cdsarc.u-strasbg.fr (130.79.128.5) or by visiting http://cdsweb.u-strasbg.fr/cgi-bin/qcat?J/A+A/, the CDS will make the classification probability of each object available in electronic form.}. Figure \ref{fig_proba} displays the resulting probability distribution. Two distinct peaks can be seen, one with probability near 0 and the other with probability near 1. These peaks match stars that the classifier can definitely identify as being MW and SMC sources, respectively. There is a flat tail with intermediate probability in between, which represents sources for which the Neural Network has more difficulties to classify. Only 537 137 stars have a probability $P$ between 0.01 and 0.9, corresponding to the 13\% of the SMC base sample.

We must establish a probability threshold $P_{cut}$ in order to acquire a classification using the probabilities that the classifier generated for each star. The star is thought to belong to the SMC if $P>P_{cut}$ and the MW if $P<P_{cut}$ (alternatively, we could deem stars with intermediate probabilities as unclassified). Fixing a low probability threshold allows us to ensure that no SMC objects are missed, but at the cost of having more "mistaken" MW stars in the SMC-classified sample. Conversely, by setting a high probability threshold, we can reduce contamination in the resultant SMC-classified sample, but at the cost of omitting some SMC stars and producing a less complete sample.

As seen in J22, a choice about the purity-completeness trade-off will determine the characteristics of the final sample and may, therefore, have an impact on the results.
To examine the impact of this trade-off, we defined two different samples in this work: 

\begin{enumerate}
    \item Complete sample ($P_{cut} = 0.01$). In this case, a cut at low probability prioritizes completeness at the cost of larger MW contamination. We determined the cut value by looking at the classification's probability histogram (Figure \ref{fig_proba}) and selecting the upper limit of the peak of small probability values.
    
    \item Optimal sample ($P_{cut}=0.31$). The probability cut in this instance was determined to be the best possible in terms of classification; the value corresponds to the ``elbow'' of the ROC curve (Figure \ref{fig_roc_precission_recall}), which is in principle the ideal compromise between completeness and purity. 

\end{enumerate}

\begin{figure*}
    \centering
    \includegraphics[width=1\textwidth]{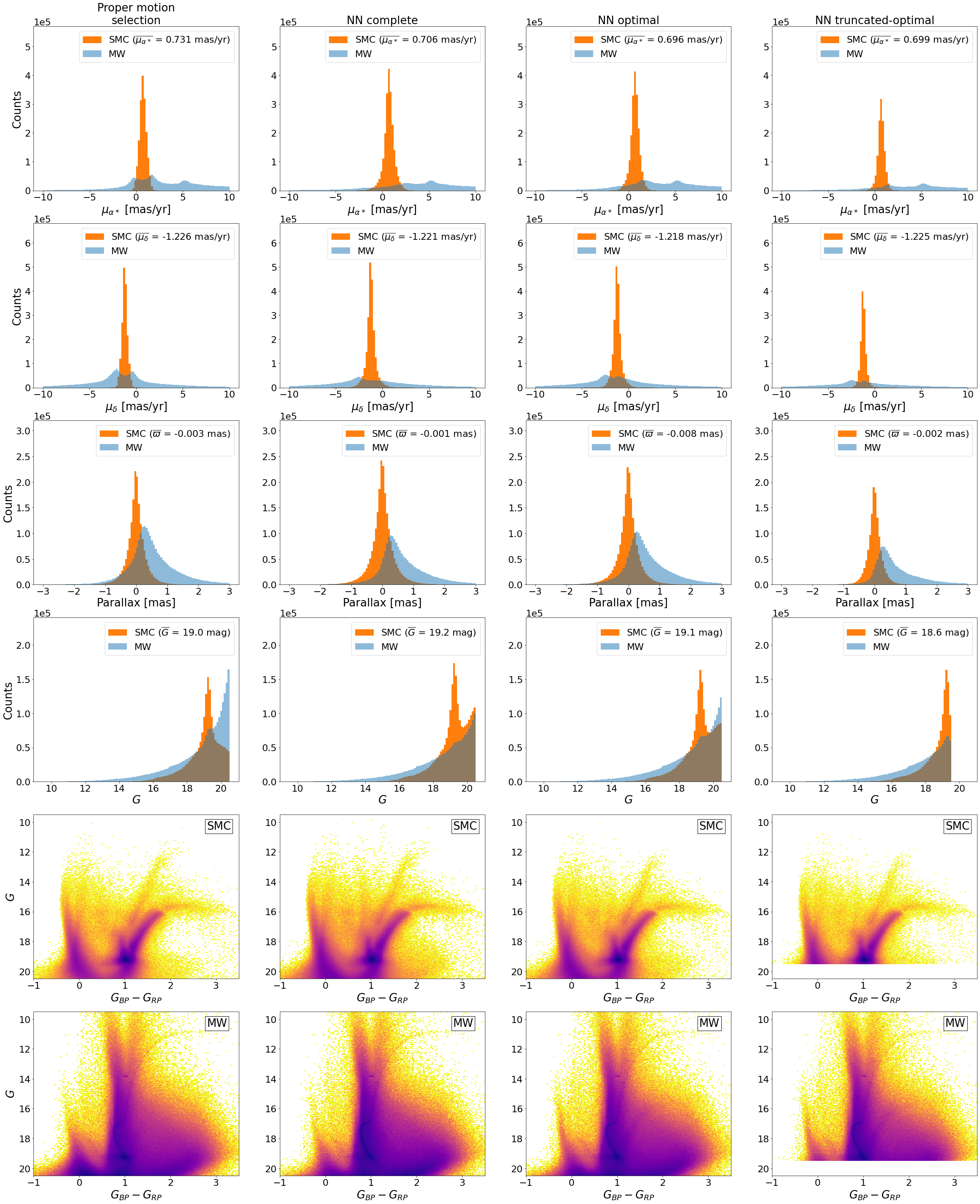}
    \caption{Astrometric and photometric characteristics of the SMC and MW samples. From left to right: PM sample, NN complete, NN optimal and NN truncated-optimal samples. In the first four rows, we show distributions of proper motion in right ascension and declination, parallax, and G magnitude, respectively, of the SMC (orange) and MW (blue) samples. In the last two rows, we show the colour-magnitude diagram of the samples classified as SMC and MW, respectively. Color represents the relative stellar density, with darker colors meaning higher densities.}
    \label{fig_histogrames_classificador}
\end{figure*}

Additionally, and because MW stars exponentially rise at fainter magnitudes whereas SMC stars rapidly decrease beyond $G \simeq 19.5$ (see discussion in the next section), we introduced the third case after carefully studying the results for the optimal sample. We refer to it as the truncated-optimal sample ($P_{cut}=0.31$) with $G < 19.5$ mag. As mentioned above, this cut avoids a region in the faint end where the SMC training sample is not representative; by removing these stars, the MW contamination can be reduced and the stars with larger uncertainties are also discarded. Given the purity of the SMC diagrams in Figure~\ref{fig_histogrames_classificador}, we decided against making a second selection by excluding areas of the CMD diagram where contamination is more likely. 

Finally, we take into account two datasets for each of the four samples. First, the full sample, where we assume that there is no information on the line-of-sight velocities for any of the stars. Secondly, a subset of the first sample that only contains stars with \gaianospace~DR3 line-of-sight velocities is kept. These samples are referred to as the corresponding $V_{los}$ samples. In Table \ref{table_astrometry}, the second and third columns show the number of stars for each data set together with the mean astrometric information.

\begin{table*}
\centering
\begin{tabular}{|l|c|c|c|c|c|c|c|c|}
\hline 
SMC sample                &  N  & $N_{vlos}$ & $\overline{\varpi}$ & $\sigma_\varpi$ &   $\overline{\mu_{\alpha*}}$ & $\sigma_{\mu_{\alpha*}}$ & $\overline{\mu_{\delta}}$ & $\sigma_{\mu_{\delta}}$ \\ \hline
Proper motion selection & 1 720 856 & 4 014 & -0.0029        & 0.323          & 0.731                & 0.370                   & -1.226               & 0.297                  \\
NN complete      & 2 172 427    & 4 195 & -0.0013            & 0.417             & 0.706                   & 0.580                      & -1.221                  & 0.558                     \\
NN optimal     & 1 979 603    & 3 335 & -0.0083            & 0.381             & 0.696                   & 0.485                      & -1.218                  & 0.463                     \\
NN truncated-optimal     & 1 265 824    & 3 335 & -0.0018            & 0.254             & 0.700                   & 0.383                      & -1.225                  & 0.349                    \\\hline
\end{tabular}
\caption{Comparison of the SMC samples number of sources and mean astrometry between the proper motion selection (MC21) and the neural networks. Parallax is in $\mathrm{mas}$ and proper motions in $\mathrm{mas}\,\mathrm{yr}^{-1}$.}
\label{table_astrometry}
\end{table*}

\subsection{Comparison of classifications}

\begin{figure}
    \centering
    \includegraphics[width=0.50\textwidth]{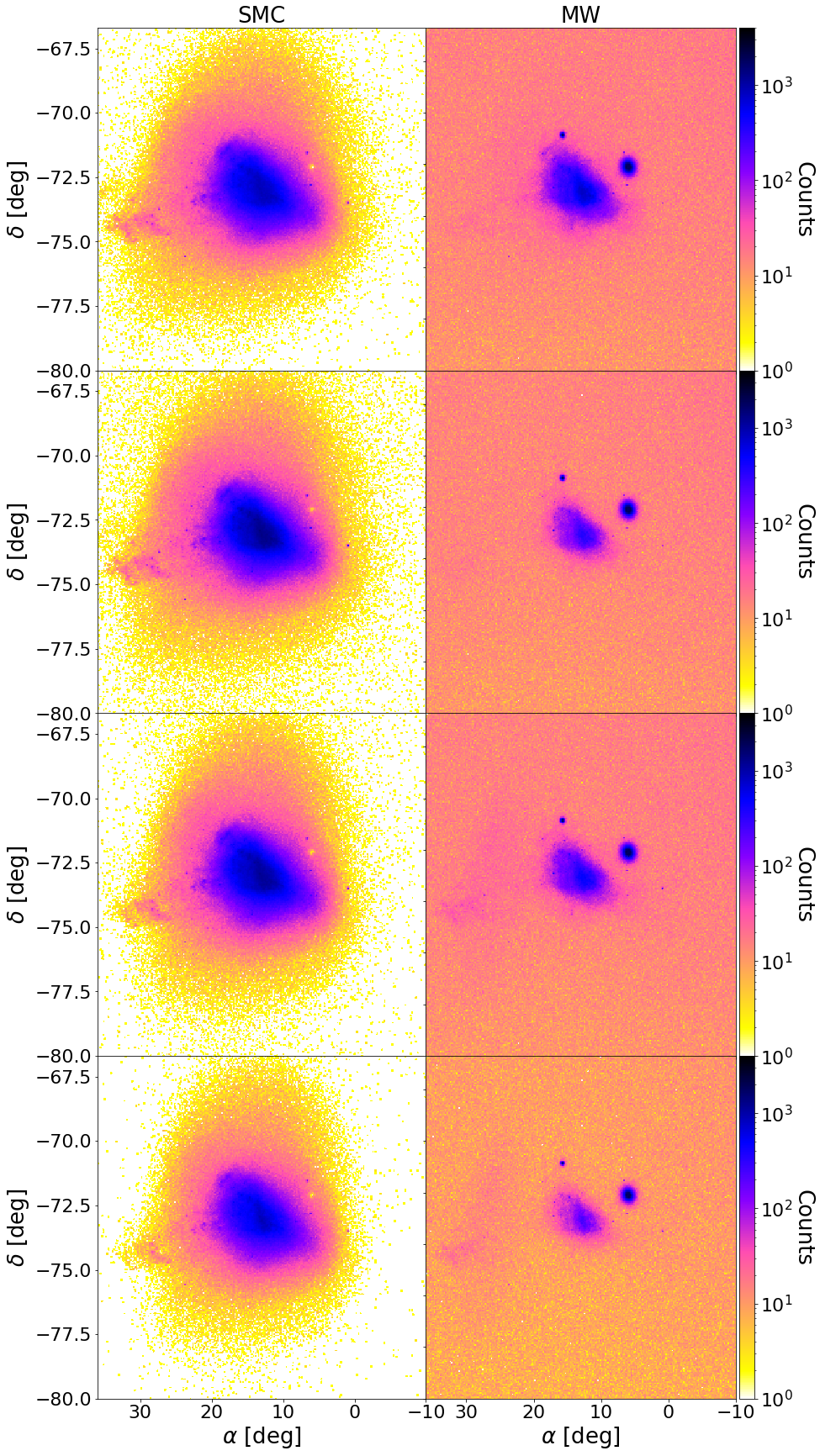}
    \caption{Sky density distribution in equatorial coordinates of both the SMC (left) and MW (right) sample obtained from the different classifiers. First row: proper motion selection classification. Second row: Complete NN classification. Third row: Optimal NN classification. Fourth row: Truncated-optimal NN classification. Note:\ in the fourth row, we display a cut in magnitude $G > 19.5$ for both the SMC and MW samples and, therefore, the total number of stars is reduced.}
    \label{fig_comparison_spatial}
\end{figure}

Figure~\ref{fig_comparison_spatial} displays the sky density distributions for the classified SMC/MW members in our various samples. We provide the SMC selection for each sample in the left column, and the sources designated as MW are displayed in the right column. Proper motion selection is the first row, followed by the three NN-based selection strategies, and each row corresponds to one selection technique. As may be expected, the outcomes of the proper motion-based selection closely resemble those of MC21. 

Since an anomalous classification in the SMC outskirts is not seen in these figures, we notice that the restricted spatial distribution of the SMC training sample (square region in top-left panel of Figure \ref{fig:SMCbasesample}) does not pose an issue for extrapolating the membership outside this region.

Additionally, we observe that sources identified as MW by all four samples exhibit an overdensity in the SMC central part, the most populated region, indicating that SMC stars were misidentified. Two globular clusters Tuc 47 and NGC362 are successfully removed from the SMC samples, see the concentration of stars around $(\alpha, \delta) \simeq (5^\circ, -72^\circ)$ and ($16^\circ,-71^\circ$), respectively. Moreover, we observe that, in accordance with the concept of the probability cut, fewer stars are categorized as belonging to the MW the more complete the SMC sample is. In this regard, a cross-match between the complete sample and the proper motion selection sample reveals that the latter almost entirely contains the former: of the 1~720~856 stars in the proper motion sample, 1~697~614 of them are included in the complete sample, and the complete sample also contains nearly four hundred thousand additional stars. Regarding the MW samples, we can estimate their SMC contamination by comparing its density with the one of an uniform sky field observed nearby, but away from the SMC centre; the observed overdensity gives an estimation of the "excess" of SMC stars. From this comparison the percentage of SMC stars in the MW sample is estimated to be around 5-10\%, being the less contaminated one the MW optimal sample.

We also notice that the astrometric parameter dispersion decreases from the NN complete to the NN truncated-optimal samples. This is to be expected given that the samples' distance and velocities are more similar due to the stricter sequence of selection criteria.

In Figure \ref{fig_histogrames_classificador}, we compare the astrometry and photometry distribution of the different SMC samples. In the proper motion selection sample, the distribution of proper motion is observed to be narrow around the bulk motion of the SMC due to the severe cut in proper motion enforced, however in the MW classification, two minor peaks are evident after the SMC. The NN samples do not reveal this misclassification. We observe a secondary peak in the right ascension proper motion around $5.2$ $\mathrm{mas}\,\mathrm{yr}^{-1}$ which corresponds to the systemic motion of Tuc47 \citep{helmi18}.
The truncated-optimal sample has the narrowest parallax distribution among the four LMC samples, which are all quite similar to one another. The $G$ magnitude distributions in the four SMC selections vary significantly from one another. Both the PM and the NN samples have a $G$ magnitude peak at $G \sim 19$ mag, which is related to the SMC stars, and a secondary peak at the limiting magnitude $G = 20.5$ mag, which corresponds to the MW contamination. Due to this, we define the truncated-optimal sample by subtracting the secondary peak from the optimal sample, as mentioned above. This secondary peak is caused by the exponential distribution in $G$ of the MW stars, arising from the logarithmic relation between the stellar flux and the apparent magnitude combined with the magnitude cut and the spatial distribution of the stars in the disk. The SMC stars, on the other hand, exhibit a significant peak at $G \simeq 19$ mag, slightly differing between samples depending on the amount of MW misclassified sources.

All SMC samples have a fairly similar CMD. Only minor variations are visible in the MW selection of the optimal and truncated-optimal samples, which comprise, as expected, sources of the red giant branch of the SMC that the NN classifier misidentifies as MW.

\section{External validation of the classification}
\label{sec:NNvalidation}

\begin{figure}
    \centering
    \includegraphics[width=0.50\textwidth]{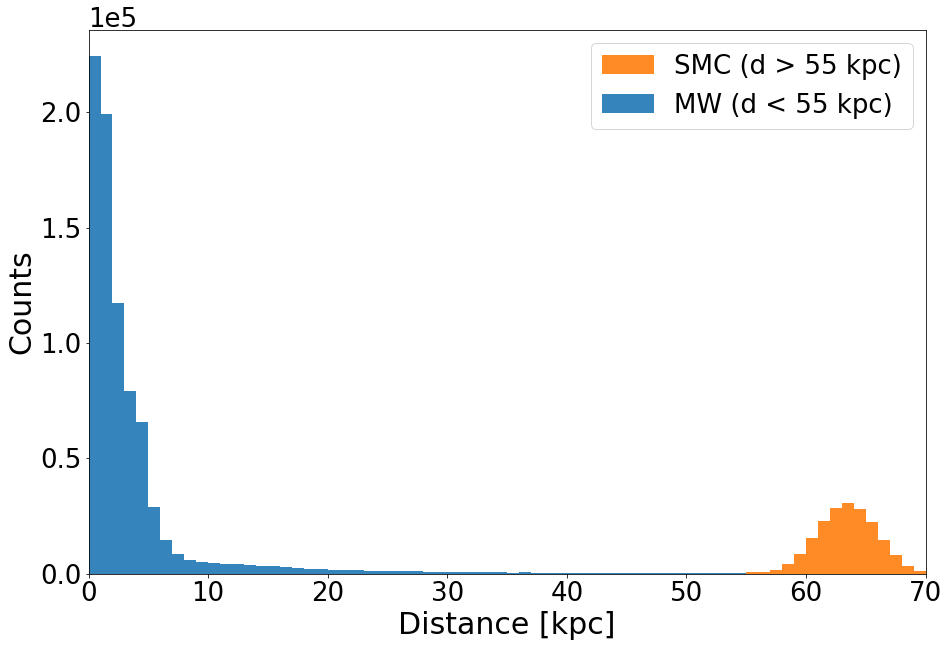}
    \caption{StarHorse validation sample distance distribution. In blue (orange), the StarHorse stars classified as MW (SMC) according to the $d = 55$ kpc criteria.}
    \label{fig_SH}
\end{figure}

In order to validate the results of our selection criteria we compare each of the generated samples with external independent classifications. To do so, we cross-matched our samples with dedicated catalogues of the SMC chosen to have a high degree of purity in the visible band. For this reason, we exclude from this exercise the VMC survey \citep{cioni11} for being in the near-infrared and the SMASH survey \citep{Nidever2017} for not performing any contamination study, and we use the following:

\begin{itemize}
    \item SMC Cepheids \citep{ripepi17}: we used the 4~793 Cepheids from the paper's sample as a set of highly reliable SMC objects. Using a 0.3" search radius to find high confidence matches and keeping 4~788 stars, we cross-matched the positions supplied in the study with the \gaia DR3 catalogue to obtain the \gaia DR3 data. To make a final selection of 4~765 SMC Cepheids, we introduced a cut with a 10$^{\circ}$ radius around the SMC center (replicating our base sample).
    \item SMC RR-Lyrae \citep{muraveva18}: we employed the 2~997 RR-Lyrae sample from the paper as high-reliability SMC objects in a manner similar to the foregoing. After the sample is cross-matched with the \gaia DR3 catalogue, it is downsized to 2~982 stars, and then we cut a final sample of 2~922 SMC RR-Lyrae in a 10$^{\circ}$ radius around the SMC center.
    \item StarHorse \citep{anders22}: using a cut of 10$^\circ$ around the SMC center, we cross-matched this catalog with the \gaia DR3 data and obtained a sample of 1~000~066 stars. We distinguished MW and SMC stars using the StarHorse distances, but with a cutoff of $d = 55$ kpc, using criteria similar to those put forward in \citet{Schmidt20,schmidt22} for the LMC. This choice is supported by the StarHorse sample's distance distribution, which is depicted in Figure~\ref{fig_SH}. A very stringent categorization is produced by a cut in $d = 55$ kpc, reducing the pollution of MW stars (see discussion below). As a result, we are left with a StarHorse SMC sample of 193~402 stars and a StarHorse MW sample of 806~660 stars. Notice that this sample only has stars up to $G=18.5$. 
\end{itemize}

The Cepheids and RR-Lyrae datasets contain objects that are highly reliably identified as SMC stars, therefore they are used to assess how complete our classification of SMC objects is (i.e., how many we lose). On the other hand, because the StarHorse classification is imperfect, this sample can be used to estimate the contamination brought on by incorrectly identified MW stars. Furthermore, the estimated amount of MW contamination in the classification will be a ``worst case'' scenario because of the extremely strict criteria utilized in StarHorse for the separation (cut in $d = 55$ kpc). 

Table~\ref{table_validation} compares the outcomes of our four classification criteria as they were applied to the stars in the three validation samples. The results using the Cepheids, RR-Lyrae, and StarHorse SMC validation samples reveal that the completeness of the resulting SMC classifications is excellent, typically exceeding $95$\%. The truncated-optimal sample is the exception, where the cut in faint stars reduces the RR-Lyrae's completeness. 

\begin{table*}
\centering
\begin{tabular}{|l|c|c|c|c|}
\hline
Stars classified as SMC         & \begin{tabular}[c]{@{}c@{}}SMC Cepheids\\ (4~765)\end{tabular} & \begin{tabular}[c]{@{}c@{}}SMC RR-Lyrae\\ (2~922)\end{tabular} & \begin{tabular}[c]{@{}c@{}}SMC StarHorse\\ (193~402)\end{tabular} & \begin{tabular}[c]{@{}c@{}}MW StarHorse\\ (806~664)\end{tabular} \\ \hline
Proper motion selection         & 4 578 (96.1\%)                                                   & 2 447 (83.7\%)                                                   & 190 166 (98.3\%)                                                    & 114 354 (14.2\%)                                                   \\
NN complete       & 4 688 (98.4\%)                                                   & 2 814 (96.3\%)                                                   & 191 692 (99.1\%)                                                    & 125 200 (15.5\%)                                                   \\
NN optimal           & 4 599 (96.5\%)                                                   & 2 694 (92.2\%)                                                   & 186 063 (96.2\%)                                                    & 110 704 (13.7\%)                                                   \\
NN truncated-optimal & 4 598 (96.5\%)                                                   & ~~ 821 (28.1\%)                                                   & 186 063 (96.2\%)                                                    & 110 704 (13.7\%)                                                   \\ \hline
\end{tabular}
\caption{Matches of the classified SMC members in our four considered samples against the validation samples. The total number of stars, which is listed beneath the sample name, is used to determine percentages.}
\label{table_validation}
\end{table*}

\begin{figure}
    \centering
    \includegraphics[width=0.50\textwidth]{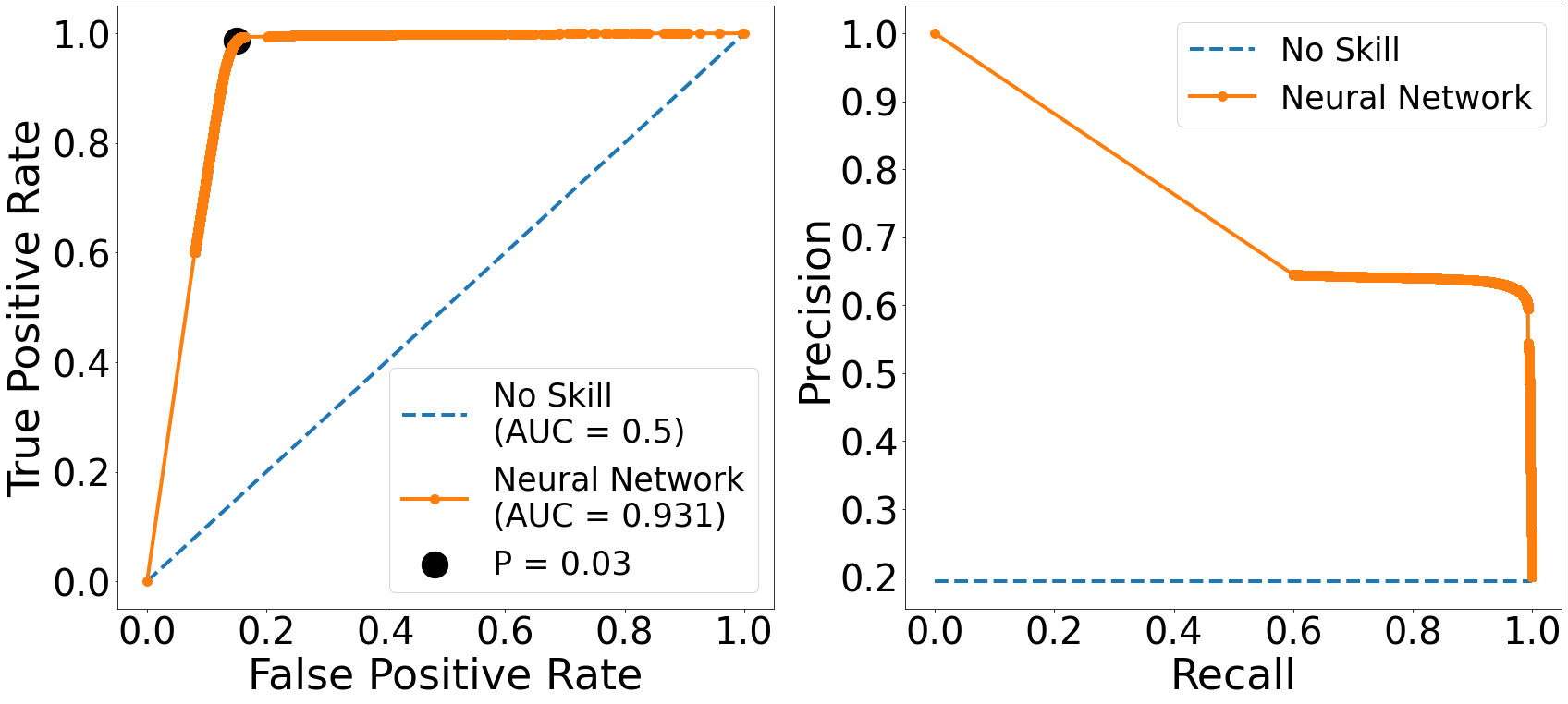}
    \caption{Evaluation metrics for the Neural Network classifier performance using the StarHorse sample. Left: ROC curve.  Black dot is in the ``elbow'' of the ROC curve and it shows the best balance between completeness and purity. Right: Precision-Recall curve. In both cases, we compare our model (orange solid curve) with a classifier that has no class separation capacity (blue dashed curve).}
    \label{fig_roc_precission_recall_SH}
\end{figure}

On the other hand, the relative contamination by MW stars it is more challenging to evaluate in the samples. We depend on an external comparison, the StarHorse distance-based classification, with the caveat that this classification also includes its own classification errors. In order to do this, we recalculate the Precision-Recall curve, using the StarHorse classification as a reference this time; the outcome is depicted in Figure \ref{fig_roc_precission_recall_SH}. We can observe that the precision essentially stays flat across the entire plot's range, or across the entire range of probability threshold values. This suggests that the complete and optimal samples both have identical relative contamination since the more restrictive we are, the more MW stars we remove, but also we lose more SMC stars. According to the precision values in Figure \ref{fig_roc_precission_recall_SH}, using the classification based on StarHorse distances as a reference, the relative contamination of our samples could be around 40\%; this is a worst-case scenario because we used a very restrictive distance cut. These statistics need to be interpreted carefully because the MW-SMC separation based on StarHorse distances is not a perfect classification criterion and actually uses less data than our criterion. Although many stars still have intermediate distances that fall between the Magellanic Clouds and the MW as a result of the multimodal posterior distance distributions, these populations are plainly evident as overdensities in the maps as mentioned in the StarHorse publication \citep{anders22}.

\begin{figure*}
    \centering
    \includegraphics[width=1\textwidth]{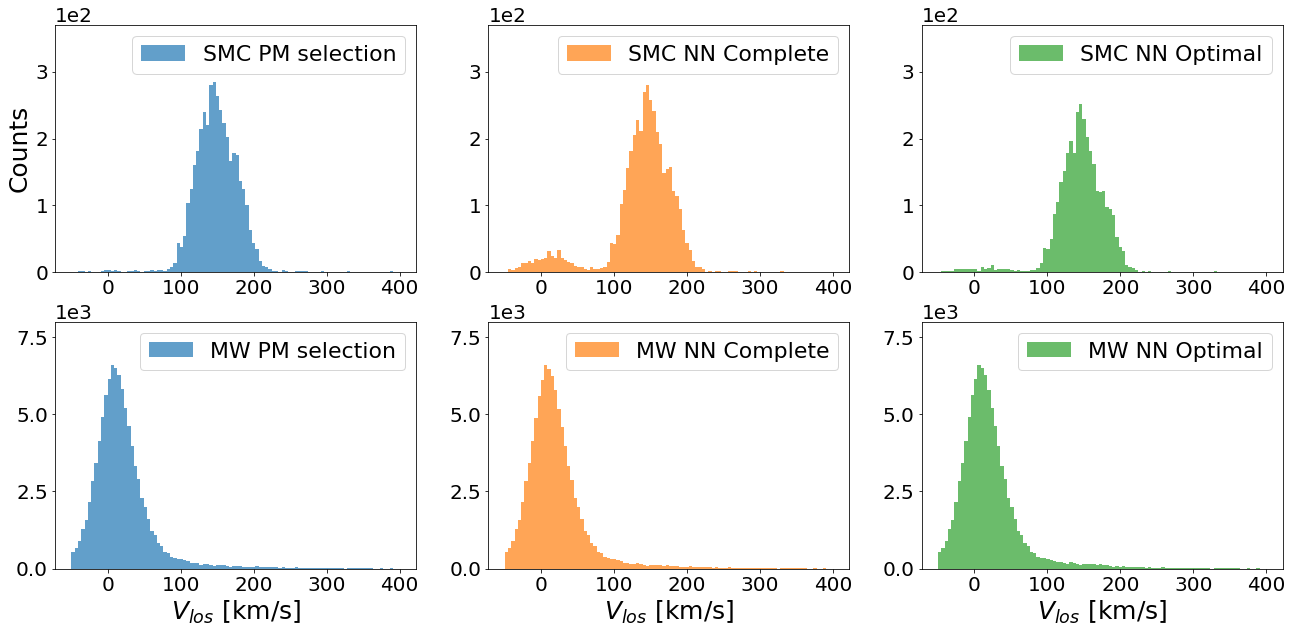}
    \caption{Line-of-sight velocity distribution for the stars classified as SMC (top) and MW (bottom). We show the three $V_{los}$ sub-samples of the PM selection (left), NN complete (middle) and NN optimal (right) samples.}
    \label{fig:vlos_histogram}
\end{figure*}

These findings indicate that there may be a few tens of percent of MW stars in our samples, but we can further investigate using the line-of-sight velocities in \gaia DR3, which are only available for a (small) subset of the full sample. These line-of-sight velocities have distinct mean values for the MW and SMC and are not used by any of our classification criteria and therefore providing an independent check. The contamination of the SMC sample is evident from the histograms of line-of-sight velocities plotted separately for MW and SMC stars in Fig.~\ref{fig:vlos_histogram}. This contamination is most likely far lower than the values mentioned above. For instance, we estimate the MW contamination to be around $10\%$ if we take into account the SMC NN complete sample and (roughly) separate the MW stars with a cut at $V_{los} < 75$ \kms. Also, this check is not entirely representative since only stars at the bright end of the sample ($G \lesssim 16$) are included in the subset of \gaia DR3 stars having observed line-of-sight velocities. 

Finally, we made a new query to the \gaia archive similar to the one described in Section \ref{sec:gaia_base_sample}. This time, we select all the sources within a 10$^\circ$ radius in a nearby area with uniform sky density from the \gaia DR3 database. By doing so, we may estimate the number of MW stars that should be present in locations that our \gaia base sample covers. We found 932~332 stars from this new query, so we may anticipate a comparable number of MW stars in the area we chose to surround the SMC. Given that the \gaia base sample contains 4~047~225 objects and the number of objects classified as SMC (Table \ref{table_astrometry}) is around 1 - 2 million, the number of stars classified as MW is around 3 - 2 million; therefore, we can conclude that our NN SMC samples prioritise purity over completeness since there are too many stars classified as MW (an excess of 1 to 2 million). This is also clear from the right panels of Figure \ref{fig_comparison_spatial}, where the pattern of SMC contamination is displayed in the distribution of stars classified as MW.

\section{Conclusions}
\label{sec:conc}

In this work, we present a new SMC/MW classification method which is compared with previous selection strategies based on the proper motion. It is based on neural networks and trained using a MW+SMC simulation created by GOG. We created two SMC samples using various probability cuts, $P_{cut}$, the NN complete, with $P_{cut}=0.01$, and the NN optimal sample, with $P_{cut}=0.32$, which corresponds to the best value according to the ROC curve. In order to remove any remaining contamination from incorrectly categorised faint stars, we added an additional cut to this final sample at the apparent $G$ magnitude of $G<19.5$ mag, creating the NN truncated-optimal sample. Moreover, we created sub-samples that contain both proper motions and line-of-sight velocities by using the recently released spectroscopic line-of-sight velocities provided in \gaia DR3. Finally, we successfully validated our classifier using external and independent classifications: SMC Cepheids, SMC RR Lyrae and SMC/MW StarHorse stars. In general, the estimated contamination of MW stars in each of the SMC samples is about $10-40\%$, being the ``best case'' for the bright stars ($G > 16$), which belong to the $V_{los}$ sub-samples, and the ``worst case'' for the full SMC sample determined by the very stringent criteria used for the separation in the StarHorse validation sample. A further check based on the comparison with a nearby area with uniform sky density indicates that the global contamination in our samples is probably close to the low end of the range, around $10\%$.

\section*{Acknowledgements}
This work has made use of data from the European Space Agency (ESA) mission {\it Gaia} (\url{https://www.cosmos.esa.int/gaia}), processed by the {\it Gaia} Data Processing and Analysis Consortium (DPAC, \url{https://www.cosmos.esa.int/web/gaia/dpac/consortium}). Funding for the DPAC has been provided by national institutions, in particular the institutions participating in the {\it Gaia} Multilateral Agreement. OJA acknowledges funding by l'Agència de Gestió d'Ajuts Universitaris i de Recerca (AGAUR) official doctoral program for the development of a R+D+i project under the FI-SDUR grant (2020 FISDU 00011). OJA, MRG, XL and EM acknowledge funding by the Spanish MICIN/AEI/10.13039/501100011033 and by "ERDF A way of making Europe" by the “European Union” through grant RTI2018-095076-B-C21, and the Institute of Cosmos Sciences University of Barcelona (ICCUB, Unidad de Excelencia ’Mar{\'\i}a de Maeztu’) through grant CEX2019-000918-M.

\bibliographystyle{aa}
\bibliography{mylmcbib} 

\label{lastpage}

\end{document}